\documentclass{eas}
\usepackage{graphicx,epsfig}
\begin{document}

\title{Stability and evolution of super-massive stars (SMS)} 
\runningtitle{Just \& Amaro-Seoane: Super-massive stars}
\author{A. Just}\address{Astronomisches Rechen-Institut, M\"onchhofstra\ss e 12--14
D-69120 Heidelberg, Germany;\\
\email{just@ari.uni-heidelberg.de\ \&\ pau@ari.uni-heidelberg.de}}
\author{P. Amaro-Seoane}\sameaddress{1}
\begin{abstract}
  Highly condensed gaseous objects with masses larger than
  \mbox{$5\cdot 10^4 M_{\odot}$} are called {\em super-massive stars}.
  They are thought to be possible precursors of super-massive black
  holes in the centres of galaxies. In the quasistationary contraction
  phase, the hydrostatic equilibrium is determined by radiation
  pressure and gravitation. The global structure is that of an $n=3$
  polytrope at the stability limit. Small relativistic corrections for
  example can initiate a free fall collapse due to the 'post
  Newtonian' instability. Since the outcome of the final collapse -- A
  super-massive black hole or hypernova -- depends sensitively on the
  structure and the size of the object, when the instability sets in,
  it is important to investigate in more detail the contraction phase
  of the SMS. If the gaseous object is embedded in a dense stellar
  system, the central star cluster, the interaction and coupling of
  both components due to dynamical friction changes the energy balance
  and evolution of the SMS dramatically.  Dynamical friction between
  stars and gas, which can be estimated semi-analytically (see Just
  \etal\ \cite{jus86}), has three different effects on the
  two-component system: 1) The gas is heated by decelerating the
  stars. This may stall the contraction process for a while until the
  stars in the 'loss cone', these which cross the SMS, lost their
  kinetic energy (for the total heating rate see Amaro-Seoane \&
  Spurzem \cite{ama01}). 2) This cooling of the loss cone stars lead
  to a mass segregation in the stellar component resulting in a much
  more condensed central stellar core. 3) The inhomogeneities due to
  the gravitational wakes in the gas changes the effective absorption
  coefficient of the gas. This affects the condition for hydrostatic
  equilibrium and may give essential deviations from the $n=3$
  polytrope. We discuss in which evolutionary stages and parameter
  range these interaction processes are relevant and how they can
  influence the stability and evolution of the SMS.
\end{abstract}
\maketitle
\section{Introduction}
In the standard picture of galactic structure, most galaxies harbour a central
Super-massive Black Hole (SMBH), which is quiet in normal galaxies and
responsible for quasar and Seyfert activities in active galaxies. But the
building process of SMBHs is still not fully understood. Is the way of creating
a SMBH with more than $10^6 M_{\odot}$ a
direct collapse of a gas cloud, is it the growth of a small seed black hole in a
dense star cluster, or
is it the remnant of a hypernova of a super massive star like object (SMS)? 

In the building process of a SMBH two basic paths can be distinguished (Rees
\cite{ree84}). At the
stellar dynamical path the gas cloud initially forms a massive central star
cluster or bulge. Stellar and dynamical evolution including star-star collisions than form a
seed black hole with initial mass up to $10^4 M_{\odot}$. Subsequently this
black hole grows by accretion of gas and stars. Here the basic problem of the
growth is the feeding rate. Only a very small fraction of stars have highly excentric orbits
to reach the tidal radius of the black hole and refilling this 'loss-cone' is a
complicated matter (for details see Amaro-Seoane \& Spurzem \cite{ama01} and
references therein). The gas dynamical path starts also with a collapsing gas
cloud, but here the cloud collapses directly to a highly condensed gas ball.
The lifetime of this super-massive object is very short (see section \ref{sms1}) 
such that
star formation in this condensed phase does not occur. The dynamical collapse can go on to end up
with a SMBH, or it can reach a hydrostatic state, which than contracts by
radiative cooling. We call a gas ball with mass $M>5\cdot 10^4 M_{\odot}$ in
hydrostatic equilibrium a super-massive
star (SMS). It is not proofed that a direct collapse is possible, because
explosive hydrogen burning would destroy more or less the whole object. Even the
final collapse of a
SMS, which becomes unstable due to the post-Newtonian instability, will be
stopped by hydrogen burning (Fuller \etal\ \cite{ful86}).

In a more realistic scenario of galaxy formation, a central bulge or dense star
cluster will form first, and at a later stage there will occur strong gas infall
initiated by a galaxy merging process or a bar instability. In this case the
super-massive gaseous object will not be isolated and interaction processes with
the star cluster can change the evolution dramatically. Hara (\cite{har78})
showed in an analytic approximation, that heating of the SMS by drag forces of
the stars crossing the SMS can stop the contraction for some time. In a
numerical simulation Langbein \etal\
(\cite{lan90}) showed that the energy input due to star-star collisions can
dominate over heating by drag forces and leads to a rebounce
of the SMS. 

In the next section we give an overview over
the properties of a SMS and in section \ref{sms2} we discuss the interaction
processes of the SMS with the surrounding star cluster in a more complete way.

\section{The standard SMS \label{sms1}}
\begin{figure}[!ht]
\begin{center}
\epsfclipon
{\epsfig{file=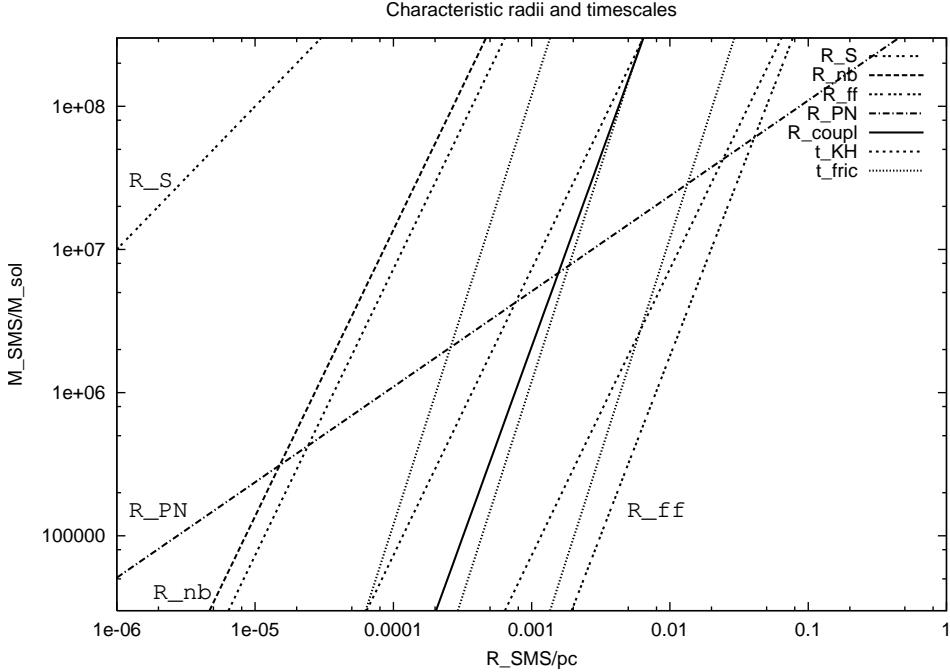,height=\textwidth,angle=270}}
\end{center}
\caption[]{
The regime of hydrostatic contraction of a SMS bounded by $R_{ff}$, $R_{nb}$,
and $R_{PN}$ is shown. For comparison the Schwarzschild radius $R_S$ is also
plotted. Lines of constant contraction time
($t_{KH}=10^2,\,10^3,\,10^4yr$ from right to left) and of constant friction
timescales ($t_{fric}=10^1,\,10^3,\,10^5yr$ from left to right) show that
the efficiency of dynamical friction grows quickly during contraction. The regime,
where the friction
timescale is shorter than the contraction time, $R<R_{coupl}$, is also shown. For
the stars a typical velocity of 5000 km/s is used. For the friction timescale we
assumed that the stars move only 1\% of their orbital time through the SMS due
to the high excentricity. Stars on more circular orbits are decelerated faster.
}
\label{fig1}
\end{figure}

In a SMS radiation pressure $P_{rad}$ dominates strongly over thermal
pressure $P_g$, the fraction of which is
\begin{equation}
\beta = \frac{P_g}{P_g+P_{rad}} = 7.8\cdot 10^{-4} M_8^{-1/2}
\end{equation}
where $M_8$ is the mass in units of $10^8 M_{\odot}$. $\beta$ is independent of
radius
($R_{pc}$ given in $pc$) and does not change during contraction of the SMS. 
For $\beta<<1$  
the structure of the SMS is very near to a n=3 polytrope, which is the
stability limit with zero total energy. The SMS is determined by mass $M$ and
radius $R$. The central temperature $T_c$, central density $\rho_c$,
gravitational and total energy ($E_g$ and $E_{tot}$) are given by
\begin{eqnarray}
T_c &=& 1.9 \cdot 10^{4} K \,M_8^{1/2} R_{pc}^{-1} \\
\rho_c &=& 54\overline{\rho} = 10^{-13} \frac{g}{cm^3}\, M_8 R_{pc}^{-1} 
         = 1.5 \cdot 10^{9} \frac{M_{\odot}}{pc^3}\, M_8 R_{pc}^{-1}\\
E_g &=& -\frac{3}{2}\frac{GM^2}{R} = -1.3\cdot 10^{57} erg \,M_8^2 R_{pc}^{-1}\\
E_{tot} &=& \frac{3}{2}\beta E_g = -1.5\cdot 10^{54} erg \,M_8^{3/2} R_{pc}^{-1}
\quad ,
\end{eqnarray}
respectively. The contraction rate of the SMS is determined by the Eddington
luminosity
\begin{equation}
L_{Ed} = 1.3\cdot 10^{46} \frac{erg}{s}\, M_8
\end{equation}
leading to a contraction time
\begin{equation}
t_{KH} = \frac{-E_{tot}}{L_{Ed}} = 3.7 yr \, M_8^{1/2} R_{pc}^{-1}
\end{equation}
In parameter space (the $R,M$-plane), the regime of hydrostatic, stable equilibrium
(where a SMS can exist) is bounded at the high radius end (at $R_{ff}$) where the
Kelvin-Helmholtz time $t_{KH}$ (contraction time) equals the free fall time.
At the
low radius
end it is bounded by the post-Newtonian instability ($R_{PN}$) or for lower
masses when nuclear
burning at the centre sets in ($R_{nb}$). In figure \ref{fig1} the relevant
regime for a SMS is shown. The Schwarzschild radius $R_S$ is far below the
 living
region of a SMS. The evolutionary tracks are horizontal lines from
right to left. The contraction time is also plotted showing that without
re-heating the lifetime of the SMS
is at most $2\cdot 10^4 yr$ before nuclear burning or the 
post-Newtonian collapse sets in.

\section{SMS in a central star cluster \label{sms2}}

In the central region of young galaxies with SMS both, stellar and gas
densities, are very high exceeding $10^8 M_{\odot}pc^{-3}$. Therefore the
interaction rate of stars is very high and drag forces of 'loss-cone' stars
(this are the stars crossing the SMS) can drive the evolution considerably.
Earlier investigations (Hara \cite{har78}, Langbein \etal\ \cite{lan90}) show
that the timescale for dynamical coupling of the stellar system with the gas
 and
for heating by stellar collisions is comparable or even shorter than the
lifetime of a SMS. 

Here we concentrate on the star-gas interaction and discuss the effects on the
structure and evolution of the SMS. Stars moving through the gas produce
density and velocity perturbations by gravitational forces and the geometrical
cross section. The consequences of these local perturbations on the global
behaviour of the SMS can be consistently estimated using a quasi-linear
perturbation theory. In this method, all physical quantities are decomposed
 into a local mean
value and a perturbation ($\rho = \rho_0 + \rho_1$ for the gas density e.g.).
Then the time dependent perturbations are computed from the set of linearised
equations and finally the quasistationary fluctuation field of the nonlinear
quantities as functions of the mean quantities are derived by taking the
statistical average over all stars. Dynamical friction
between stars and gas is given by $\langle\rho_1\nabla\phi_1\rangle$ with
gravitational potential $\phi$ (see Just \etal\ \cite{jus86}). The level of
approximation is similar to that of using Chandrasekhars dynamical friction
formula. The advantage of the fluctuation theory is that the turbulent velocity
field, corrections to the equation of state and to radiation transfer can all
 be
derived consistently. It is also straightforward to include all relevant
physical processes in the gas like radiation pressure, heating and cooling
processes, viscosity or even
magnetic fields. 

The structure of the SMS is determined by hydrostatic equilibrium and a
stationary radiation field in the diffusion limit. These assumptions are not
automatically valid for the perturbations, because here the length and
timescales are much shorter. The perturbations are time dependent and it depends
on the scale, whether the radiation field will be compressed also.
The character of the perturbations will be between the two extremes:
\begin{description}
\item[Isothermal] In small scale fluctuations the inhomogeneities of the 
radiation field diffuse quickly away. Since the temperature is determined by the
radiation, the compression of the gas
is isothermal and the sound speed relevant for determining the Mach number of
the stellar motion is the usual isothermal sound speed ($c_s\approx 100km/s$ for
a million solar mass SMS with radius $R=0.01pc$). Most stars are highly
 supersonic in this case.
\item[Adiabatic] Larger fluctuations, where the radiation field is also
compressed, are adiabatic. The Mach number is a factor of $\beta^{1/2}$ smaller 
than in the isothermal case. It is determined by the total pressure
leading to
typical Mach numbers of order unity for the stars.
\end{description}
The resultant density and velocity fluctuations of the gas and the variations in
the radiation field change the mean local properties of the SMS systematically.
In the SMS basically three different effects of the fluctuations can
be important:
\begin{description}
\item[Dynamical heating] The energy loss of the stars by dynamical friction
heats the SMS in the inner region. The heating rate depends on the relative
masses and densities of the star cluster and the SMS, and of the filling factor
of the loss cone, which is a complicated matter to estimate. If the heating rate
exceeds the cooling of the SMS, then the contraction can be stalled for a while. 
\item[Mass segregation] Dynamical friction leads also to mass segregation of the
stars. Loss cone stars become confined to the SMS. This happens, if the friction
timescale $t_{fric}$ becomes smaller than the contraction time $t_{KH}$. 
In figure
\ref{fig1} an estimation of $t_{fric}$ is shown and also the coupling radius
$R_{couple}$ of the SMS
for stars with an initial velocity of 5000 km/s. In more compact SMS loss-cone 
stars are
effectively stopped by dynamical friction.
Since dynamical friction
has its maximum just above the sound speed, confined stars will be decelerated
quickly below the sound speed building a new highly condensed stellar core. Then
the question is, whether this new core can decouple from the surrounding to
build a new smaller SMS-star cluster system with higher density. 
\item[Equation of state] The turbulent velocity field and non-isothermal density
fluctuations change the relative contribution of
radiation pressure. This results in a
different mean $\overline{\beta}$ of
the form
\begin{equation}
\overline{\beta} = \beta_0(1+C\frac{\langle\rho_1^2\rangle}{\rho_0^2}) > \beta_0
\end{equation}
which stabilises the SMS. Additionally the direct influence of the
inhomogeneities on radiation transfer may generally destroy the $n=3$ behaviour
of the SMS. Then the post-Newtonian instability cannot occur.
\end{description}
As a bottom line we find, that the interaction of the stellar component and the
SMS can give rise to a secular evolution away from the unstable zero-energy
state to a more stable structure. Additionally mass segregation may lead to a
decoupling of the core region resulting in a sequence of SMS, the smaller and
more compact ones nested in the larger ones. In order to show, whether this kind
of evolution can happen, we will do detailed semi-analytical model calculations.

\end{document}